\documentclass[article]{aa}  

\doi{10.1051/0004-6361/202347789}

\usepackage{graphicx}
\usepackage{txfonts}
\usepackage[left, edtable]{lineno}

\usepackage{xcolor}
\usepackage{svg}
\usepackage{hyperref}

\begin{document}


   \title{Solar wind entry into Mercury’s magnetosphere: Simulation results for the second swingby of BepiColombo}


   \author{D. Teubenbacher\inst{1,2}, W. Exner\inst{3,4,5}, M. Feyerabend \inst{4}, Y. Narita\inst{4,1}, D. Schmid\inst{1}, G. Laky\inst{1}, S. Toepfer\inst{4}, U. Motschmann\inst{4}, P. A. Bourdin\inst{2}, H. Comişel\inst{6}
          }

   \institute{Space Research Institute Graz, Austrian Academy of Sciences, Graz, Austria\\
              \email{daniel.teubenbacher@oeaw.ac.at}
        \and
            Institute of Physics/IGAM, University of Graz, Graz, Austria
        \and
            European Space Agency, ESTEC, Noordwijk, The Netherlands
        \and
            Institut f\"ur Theoretische Physik, Technische Universit\"at Braunschweig, Braunschweig, Germany
        \and 
            Institut f\"ur Geophysik und Extraterrestrische Physik, Technische Universit\"at Braunschweig, Braunschweig, Germany
        \and
            Institute for Space Science, National Institute for Laser, Plasma and Radiation Physics, Măgurele, Romania
        }


 
  \abstract
   {We use a global 3D hybrid plasma model to investigate the interaction between Mercury's magnetosphere and the solar wind for the second BepiColombo swingby, evaluate magnetospheric regions, and study the typical energy profile of protons. }
   {The objective of this study is to gain a better understanding of solar wind entry and analyze simulated plasma data along a trajectory using BepiColombo swingby 2 conditions, with the goal of enhancing our comprehension of measurement data and potentially providing forecasts for future swingbys.}
   {To model Mercury's plasma environment, we used the hybrid code AIKEF and developed a method to extract the particle (ion) data in order to compute the proton energy spectrum along the trajectory of BepiColombo during its second Mercury swingby on June 23, 2022. We evaluate magnetopause and bow shock stand-off distances under average upstream solar wind conditions with the Interplanetary Magnetic Field (IMF) condition derived from the BepiColombo magnetic field measurements during the second Mercury swingby.}
   {We found that the magnetosheath on the quasi-perpendicular (dusk) side of the bow shock is thicker than that on the quasi-parallel (dawn) side, where a foreshock is formed. Multiple plasma populations can be extracted from our modeled energy spectra that assist in identifying magnetospheric regions. We observed protons of solar wind origin entering Mercury's magnetosphere. Their energies range from a few electron volts in the magnetosphere up to 10 keV in the magnetosheath.}
   {}

   \titlerunning{}
   \authorrunning{D. Teubenbacher et al.}
   \maketitle

%

\section{Introduction}

In certain aspects, Mercury's magnetosphere is a scaled-down version of the Earth's magnetosphere. However, the typical solar wind dynamic pressure at Mercury is about five times higher than at the Earth, and the average IMF magnitude is about four times stronger \citep{winslow2013}. Together with the weak intrinsic magnetic field of Mercury, a small and dynamic magnetosphere, compared to the Earth, is formed that interacts with the solar wind and its embedded interplanetary magnetic field (IMF) \citep{slavin1979, anderson2008, slavin2018}. In addition to a stronger dependency on the solar wind relative to the Earth, Mercury's magnetosphere also reacts faster to solar wind changes. The Dungey cycle, which describes the plasma circulation and magnetic flux transfer in the magnetosphere, takes only about two minutes, as MESSENGER observations show \citep{slavin2009, slavin2021}. The timescale is significantly shorter than the Dungey cycle at the Earth (about one hour, \cite{baker1996}), suggesting that the Hermean system is more dynamic and can only be considered quasi-stationary for a few minutes. Compared to the other terrestrial planets, Mercury is unique in (1) the lack of a significant ionosphere and (2) the existence of a large, electrically conducting core \citep{smith2012}. These features significantly impact the magnetospheric current system and induction effects \citep{jia2015, exner2018, exner2020, Ganushkina2015_AllCurrentSystems}. 

The pioneering work by \cite{ness1975} used the first in situ measurements of Mercury's magnetosphere (Mariner 10, \cite{dunne1978}) to provide initial insights. \cite{siscoe1975} and \cite{slavin1979} used this data together with observational data from the Earth to evaluate the solar wind stand-off distance at Mercury to about (1.3 - 2.1) $\mathrm{R_M}$ ($\mathrm{R_M}=2440$ km being the radius of Mercury), depending on the upstream solar wind condition. More than three decades later, MESSENGER \citep{solomon2007} observations revealed a more detailed insight and narrowed the estimate for the solar wind stand-off distance to (1.35 - 1.55) $\mathrm{R_M}$ \citep{winslow2013}. The average subsolar distance of the bow shock is (1.89 - 2.29) $\mathrm{R_M}$ \citep{slavin2009, winslow2013}. Previous studies support these results and further examine the interaction between Mercury and the solar wind, using both measurements and (numerical) models, under average solar wind conditions and varying IMF conditions (e.g., \cite{travnivcek2009, baker2013, jia2015, fatemi2018, fatemi2020}) or during extreme events, such as CMEs (e.g., \cite{exner2018,jia2019}).

The BepiColombo space mission \citep{benkhoff2010} will perform a total of six swingbys near Mercury before its final arrival at the planet in 2025. While the nominal mission does not begin before the arrival in the final science orbits, traversing Mercury's magnetosphere at various angles during the swingbys and reaching approaches as close as about 200 km gives unique science opportunities to investigate the planet's global magnetosphere. 

While there is no significant ionosphere, Mercury possesses a tenuous exosphere consisting mainly of sodium \citep{cheng1987, milillo2005}. This gas envelope is predominantly generated by multiple surface processes of the solar wind and solar radiation interaction with the planetary surface \citep{Wurz2010, Wurz2022_DriverParticleReleaseMercurySurface, Cassidy2015, Gamborino2018_PSDdistribution, exner2020}. Hence, there are two main sources for plasma in the magnetosphere: exospheric particles and solar wind particles. 

\begin{figure}[t]
   \centering
   \includegraphics[width=9cm]{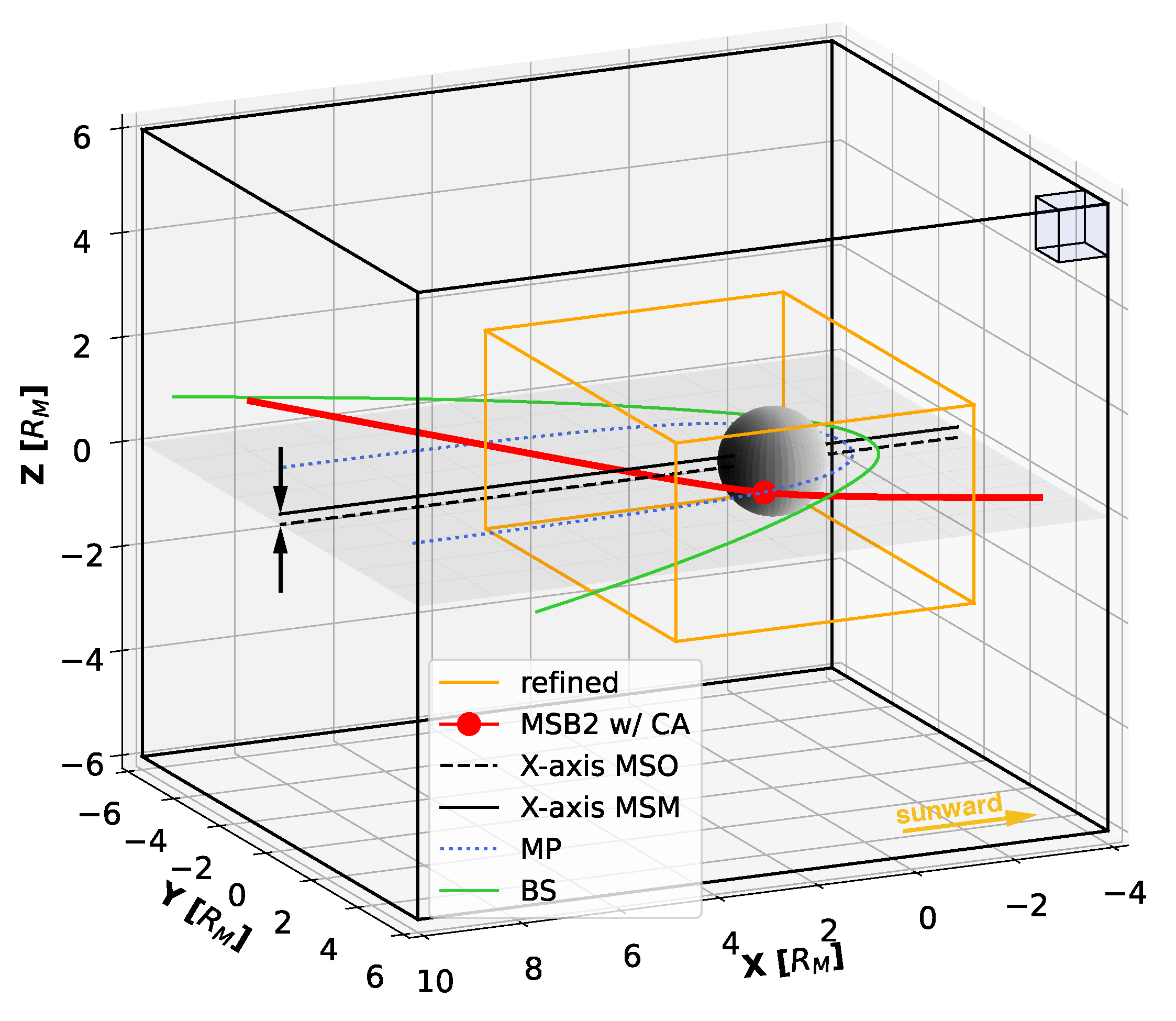}
      \caption{Overview of the AIKEF simulation box. In the top-right corner, there is an exemplary unrefined block that is further divided into cells that define the grid resolution (size ratios: block to simulation box = cell to block). The black arrows denote the difference between the MASO and MSM coordinate system, which is shifted 0.2 $\mathrm{R_M}$ in the +Z direction. The subsolar point is defined to be on the MSM X-axis; see modelled bow shock (green, solid line) and magnetopause (blue, dotted). The MSB2 trajectory with its closest approach is shown in red; the orange outline encloses the refined area (grid resolution doubled).}
         \label{fig_sim_box}
   \end{figure}

Due to limited observational coverage of previous missions, the plasma motion of planetary and solar wind ions through Mercury's equatorial low-altitude nightside magnetosphere is largely derived from scaled-down models based on the magnetosphere of Earth. However, global numerical modeling efforts have revealed particular plasma signatures, including partial plasma rings \citep{Delcourt2003, travnivcek2007, 
exner2018,Yagi2010,Yagi2017}, and increased exospheric densities \citep{exner2020} at these low altitudes. The second swingby trajectory of BepiColombo (henceforth called MSB2) is strongly equatorial, as the spacecraft passed through the southern magnetosphere and had its closest approach (CA) on the dawnward nightside at $199\,\mathrm{km}$ on June 23, 2022, obtaining measurements in the previously non-covered low altitudes (see Fig. \ref{fig_sim_box}). 

In this study, we investigate the observed particle densities and energies due to solar wind particle entry. A proton energy spectrum is computed that gives insights into regions along the trajectory where, for instance, particles are accelerated or heated. Magnetic field data from the MGF instrument \citep{baumjohann2020} on board the Mio spacecraft of BepiColombo are used to determine the outbound IMF parameters. The determined IMF direction and strength agrees well with commonly observed IMF conditions at Mercury resulting from the Parker spiral \citep{james2017}. We address two scientific questions in this study: 

How deep do solar wind particles penetrate Mercury’s magnetosphere? What do the energy spectra of the penetrating ions look like? To tackle these questions, we employed a numerical hybrid plasma approach utilizing the AIKEF model.

\section{Hybrid plasma approach}

Using a hybrid plasma model to study the interaction between the solar wind and the planet Mercury offers several advantages. The most important one in the context of this study is the advantage of a kinetic description of the ions as it enables the resolution of non-Maxwellian distributions. In addition, such models are computationally inexpensive compared to fully kinetic (particle in cell, or PIC) simulations, as the electrons are described as a fluid and the electron kinetics that happen on much smaller length and time scales than those of ions are omitted \citep{winskeomidi1996}. Therefore, physical processes down to ion kinetic time and length scales can be reasonably described using hybrid models. Hybrid modeling has been applied to investigate Mercury's plasma environment by, for example, \cite{travnivcek2007,travnivcek2009, Paral2010_HybridExosphere, Paral2013_DDA_KHI, mueller2011, mueller2012, hercik2013, fatemi2017, fatemi2018, exner2018, exner2020, Jarvinen2019_Hybrid_IonForeshock_Mercury, Aizawa2020_SHOTS_firstmodelbenchmark, lu2022}.

In this work, we used the Adaptive Ion-Kinetic Electron-Fluid code, or AIKEF, \citep{mueller2011}. It is a global 3D hybrid plasma simulation code where ions are treated kinetically and electrons are a massless charge-neutralizing fluid. A Cartesian mesh grid is used where the electromagnetic fields are computed on the nodes of the numerical mesh. While the number of particles per unit volume gives the resolution in velocity space, the resolution of the numerical mesh gives the spatial resolution. Mercury’s surface is treated as a perfect plasma absorber, which means that particles impacting the surface are removed from the simulation box. Magnetic reconnection is handled self-consistently through anomalous resistivity \citep{raeder1998, Jia2009_Ganmede_anomalousRes}. (For a detailed explanation on how the electromagnetic fields are propagated and how the AIKEF model is implemented, see \cite{mueller2011}.)

\subsection{Simulation setup}
The size of the simulation box is 12 $\mathrm{R_M}$ x 12 $\mathrm{R_M}$ x 12 $\mathrm{R_M}$ (see Fig. \ref{fig_sim_box}). We set the separation between the physical nodes to 114 km, which corresponds to 2.8 $d_{i,sw}$, with $d_{i,sw}=40$ km being the ion inertial length in the solar wind (see Tab. \ref{table:1}). However, the spatial resolution near the planet ($-2.1\,\mathrm{R_M}<X<3.9\,\mathrm{R_M}$; $-4.1\,\mathrm{R_M} < Y < 4.1\,\mathrm{R_M}$; $-1.9 \, \mathrm{R_M}< Z < 1.9\,\mathrm{R_M}$) is twice as high when using mesh refinement, resulting in a node distance of 57 km (or 1.4 $d_{i,sw}$). The model utilizes the right-handed Mercury Anti Solar Orbital (MASO) coordinate system where the X-axis is directed as the solar wind velocity (i.e., anti-sunward), the Z-axis is the positive normal to the orbital plane (anti-parallel to Mercury's dipole moment), and the Y-axis is the orthogonal complement roughly pointing toward Mercury's orbital movement direction (i.e., dawnside). To reach a quasi-steady state, the total run time of the simulation is about 760 seconds in the simulation domain, which corresponds to about 6 Dungey cycles \citep{slavin2009}. Apart from the IMF direction derived from the BepiColombo Mio MGF measurements during the outbound trajectory phase (\cite{baumjohann2020}), all other physical input parameters are adapted from the average values obtained from four years of MESSENGER observations \citep{winslow2013, anderson2012}, see Tab. \ref{table:1}.\\

\begin{table}[ht]
\caption{Input parameters used for the AIKEF model. The solar wind parameters are adapted from \cite{winslow2013}. The planetary magnetic field is considered up to the octupole moment (\cite{anderson2012}).}             
\label{table:1}      
\centering                          
\begin{tabular}{l c}        
\hline\hline                 
\textbf{Parameter} & \textbf{Value} \\    
\hline                        
    IMF direction (upstream) & $(-0.79, 0.59, 0.06)$ \\
    IMF magnitude (upstream) & $24.2 \, \mathrm{nT}$ \\
    Solar wind (H$^+$) number density & $40 \, \mathrm{particles/cm}^3$ \\
    Solar wind (upstream) bulk velocity & $400 \, \mathrm{km/s}$ \\
    Solar wind ion temperature & $17 \, \mathrm{eV}$ \\
    Solar wind ram pressure ($P_{ram}=2P_{dyn}$) & $10.6 \, \mathrm{nPa}$ \\
    Alfvén Mach number $\mathrm{M_A}$ & $4.8 \, \mathrm{}$ \\
    Planetary magnetic field (multipole) & $-190, -75, -22 \, \mathrm{nT}$ \\
\hline                                   
\end{tabular}
\end{table}

In this study, our simulations account only for protons (H\textsuperscript{+} ions) of solar wind origin as the dominant ion species. The effects of heavier solar wind ions or ions of planetary origin, such as Na\textsuperscript{+} \citep{potter1985}, are neglected. 
 
\section{Results and discussion}
\subsection{Magnetospheric regions and penetration depth}
\begin{figure*}[!p]
    \centering
        \includegraphics[width=1\textwidth]{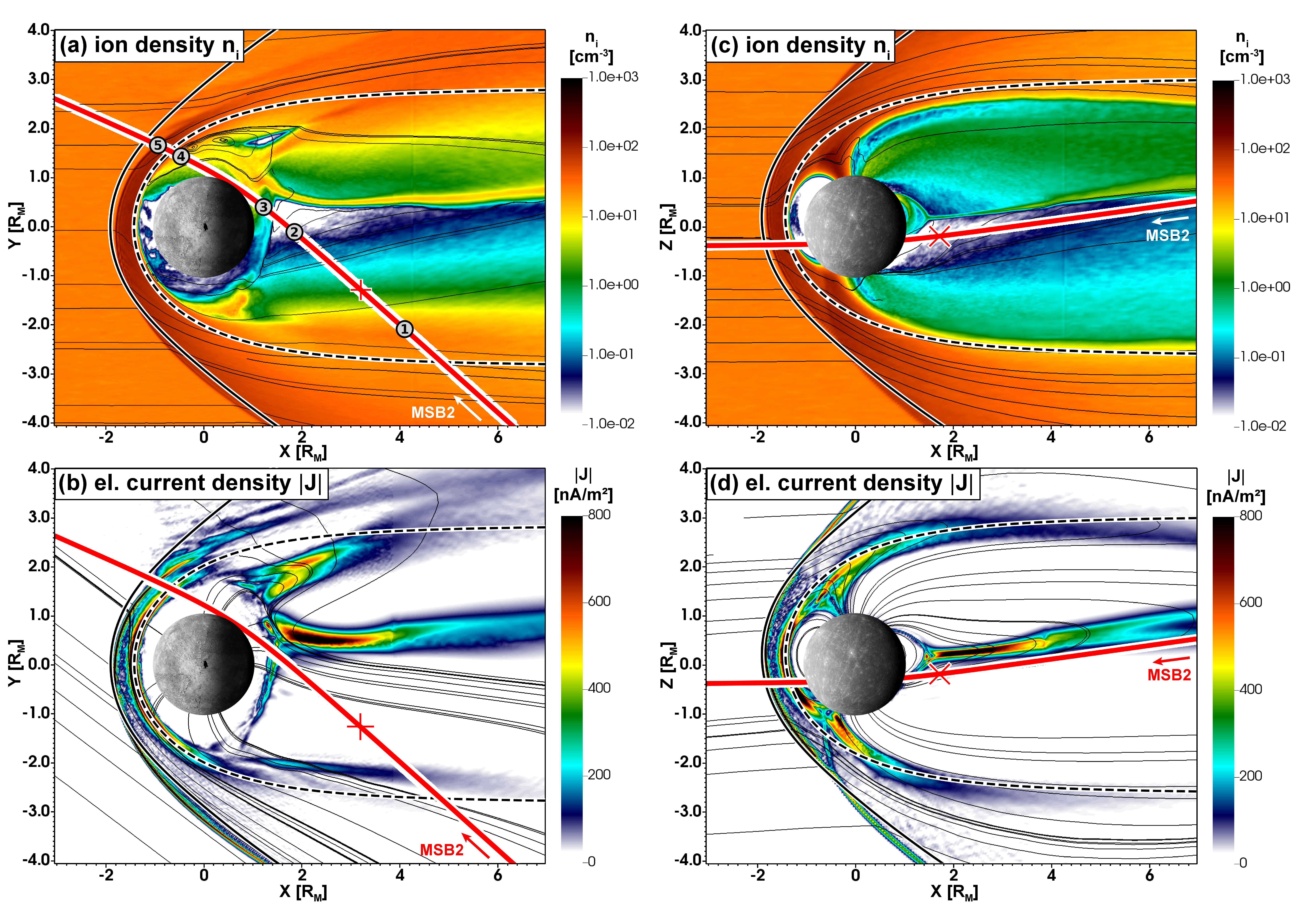}
        \caption{AIKEF MSB2 simulation overview. Panels a and b show modeled plasma density $n_i$ and electric current density $|\mathbf{J}|$ in the XY (equatorial) plane (MASO). Panels c and d illustrate the same parameters but in the XZ (meridonial) plane. The black thin lines in panels a and c represent the bulk flow velocity, whereas the black thin lines in b and d indicate magnetic field lines. The red thick lines with a white outline correspond to the projected MSB2 trajectory, with crosses indicating the points where the simulation "slice" is intersected. The models for magnetopause and bow shock are illustrated by the black dashed \citep{winslow2013} and black solid \citep{slavin2009, winslow2013} lines.}
    \label{fig:ion_density}
\end{figure*}
\begin{figure*}[!p]
   \centering
   \includegraphics[width=1\textwidth]{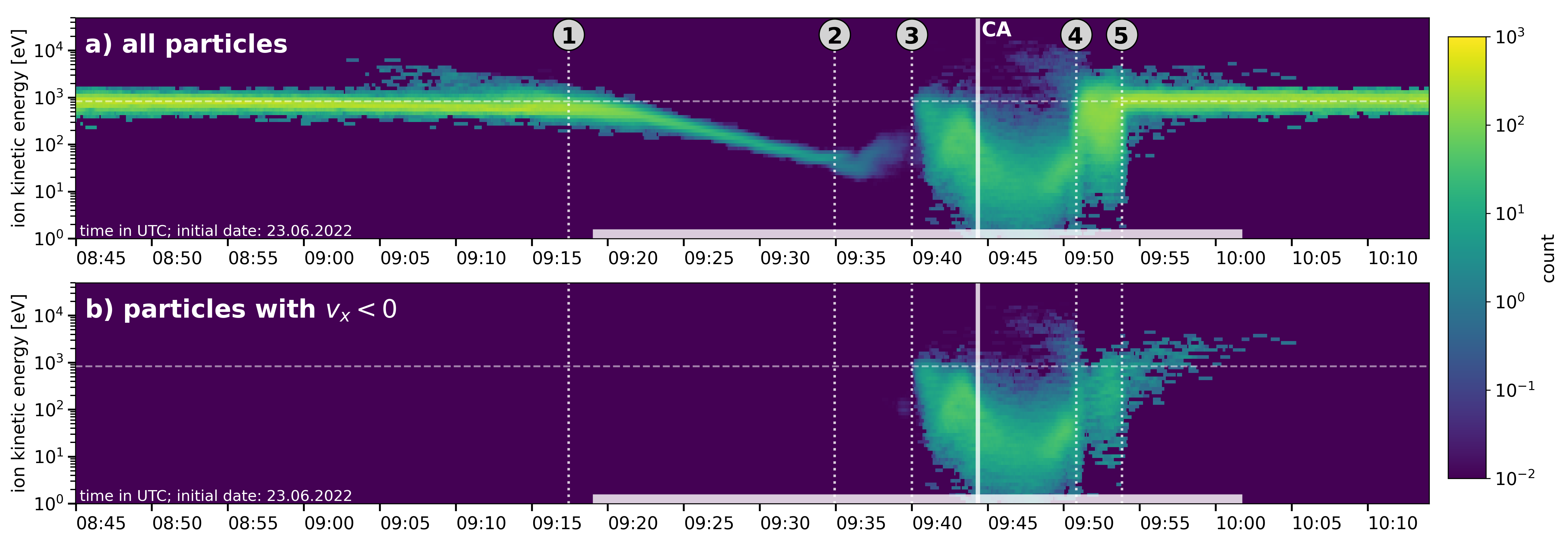}
      \caption{Model results for the proton energy spectrum along BepiColombo Mercury swingby 2 trajectory. Panel a shows all particles, and panel b only shows particles moving in a sunward ($v_x<0$) direction. The white horizontal dashed line represents the solar wind energy input parameter, and CA denotes the closest approach. The numbers 1-5 indicate magnetopsheric region or boundary crossings: (1) inbound magnetopause crossing,  (2 to 3) grazing of the magnetotail, (4) the outbound magnetopause crossing, and (5) the outbound bow shock crossing. The white bar between 09:19 UTC and 10:02 indicates the area where a higher resolution (refinement) was used.
              }
         \label{fig:ion_profile}
\end{figure*}

In Fig. \ref{fig:ion_density} panels a and c, the quasi-steady state magnetosphere of Mercury is depicted through the ion number density $n_i$ as meridional and equatorial "slices" through the simulation. Additionally, panels b and d illustrate the electric current density $|\mathbf{J}|=1/\mu_0 |\nabla \times \mathbf{B}|$, where $\mathbf{B}$ represents the magnetic field and $\mu_0$ the vacuum permeability, once again in meridional and equatorial planes. The projection of the mostly equatorial MSB2 trajectory is displayed in all subplots of Fig. \ref{fig:ion_density}. Crossings of BepiColombo through different magnetospheric regions are highlighted in panel a: (1) Inbound magnetopause crossing, between (2) and (3) grazing of the magnetotail, (4) the outbound magnetopause crossing after the closest approach, and (5) the outbound bow shock crossing at about 2 $\mathrm{R_M}$. 

We find it important to note here that Fig. \ref{fig:ion_density} displays only one slice through the simulation, and hence the MSB2 trajectory intersects each of these planes only at one point (marked with crosses). This also accounts for the tail current sheet that is tilted \citep{romanelli2022} and intersects the equatorial plane, for instance at Y$_{\mathrm{MASO}}$ = 0.5 $\mathrm{R_M}$ (see Fig. \ref{fig:ion_density} a and b). Magnetospheric regions and boundaries become more evident when considering the particle distributions in Fig. \ref{fig:ion_profile}.

In general, considering Fig. \ref{fig:ion_density} panels a and c, solar wind protons are spread throughout the magnetosphere. The lowest number densities of $>0.01 \,\mathrm{cm}^{-3}$ are found at the dayside magnetopause, and a density of about $(0.1 - 1) \,\mathrm{cm}^{-3}$ can be observed in the tail lobes (comparable to MESSENGER results \citep{zhao2020}). The density in the dayside magnetosheath is about $130 \,\mathrm{cm}^{-3}$, and the highest density of $250 \,\mathrm{cm}^{-3}$ can be seen at the cusp regions, where the particles can precipitate onto the surface, which is a driving mechanism for planetary ions that make up Mercury's exosphere \citep{raines2022, exner2020}.

Our results indicate a dawn-dusk asymmetry of the magnetosheath thickness in the equatorial plane (see Fig. \ref{fig:ion_density}, panels a and b). As the model itself is symmetric about the X-axis, the difference might be caused by the IMF direction and also influenced by the upstream solar wind condition resulting in a foreshock on the dawnside. There are several widely used models that assess the locations of the bow shock and magnetopause, including the classic Shue model \citep{shue1997}, a paraboloid model proposed by \cite{alexeev2008}, and the KT14 and KT16 models \citep{korth2015, korth2017}, among others. These models often assume (anti-)parallel IMF conditions, leading to a rotational symmetry around the X-axis. However, studies on the Earth's magnetosphere have shown clear signatures of asymmetries at the quasi-perpendicular (dusk) side of the bow shock, meaning that the angle between the bow shock normal and the IMF direction, $\theta_{bn}$, is greater than $45^{\circ}$, and the magnetosheath is thicker than on the quasi-parallel (dawn) side due to the greater extent of the shock \citep{sundberg2013, dimmock2016}. Models that include this dawn-dusk asymmetry have been developed (e.g., by \cite{tsyganenko2002}). This effect is also observable in our simulation, as can be seen in Fig. \ref{fig:ion_density} panels a and b. Compared to the modeled bow shock \citep{slavin2009} using empirical parameters from \cite{winslow2013}, a difference in the general shape as well as a slight shift to the dusk side is evident in the simulation. Moreover, while in Fig. \ref{fig:ion_density} panels a and b the quasi-perpendicular bow shock is a clear shock front, the quasi-parallel shock shows the development of a foreshock region. We find it important to note here that in Fig. \ref{fig:ion_density} the ion density is shown in the geographical equatorial plane, deviating slightly from the shifted Mercury Solar Magnetospheric (MSM)  XY-plane (see Fig. \ref{fig_sim_box}), in which the modeled bow shock and magnetopause are shown. The magnetopause seems to be symmetric around Z$_{\mathrm{MASO}}$ = 0.2 $\mathrm{R_M}$, comparing well to the magnetopause model from \cite{korth2015}.

Evaluating the areas of the highest electric current densities in Fig. \ref{fig:ion_density} panel d, the subsolar magnetopause and bow shock stand-off distances (at Z$_{\mathrm{MASO}}$ = 0.2 $\mathrm{R_M}$) are respectively $\mathrm{R_{MP}=}$ 1.45 $\mathrm{R_M}$ and $\mathrm{R_{BS}=}$ 1.85 $\mathrm{R_M}$, generally reproducing the empirical models by \cite{slavin2009} and \cite{winslow2013}. The bow shock stand-off distance is about 2\% smaller than the minimal expected value of 1.89 $\mathrm{R_M}$ \citep{winslow2013}. This might be caused by numerical inaccuracies due to the limited grid resolution in the simulation. 

At the subsolar magnetopause (most) solar wind protons can penetrate down to 1100 km toward the planetary surface and are deflected around the planet and its intrinsic magnetic field. This is illustrated by the density drop toward the planet and by the thin black lines depicting the ion bulk flow velocity in Fig. \ref{fig:ion_density} panels a and c. The penetration depth decreases slightly toward the dawnside direction, as can be seen in Fig. \ref{fig:ion_density} panel a. In addition, one can see trapped particles in terms of closed streamlines at $X=0 \,\mathrm{R_M}$, $Y=1.5 \,\mathrm{R_M}$ in Fig. \ref{fig:ion_density} panel a. As the MSB2 trajectory passes right through this area, these particles are also seen in the energy spectrum in Fig. \ref{fig:ion_profile} and are discussed in the following section.


\subsection{Penetrating ion energy profile}
To predict the ion energy profile along the spacecraft trajectory through the magnetosphere, we used a method that extracts the particle data from the completed simulation snapshot. This approach offers the flexibility to adjust the trajectory with minimal computational effort, regardless of the complexity of the underlying simulation.

To determine the plasma energy distribution along MSB2, we extracted all particles within a constant distance of 5 ion inertial lengths (in the solar wind) and gathered their velocity distribution. Subsequently, an omnidirectional velocity distribution and thus energy spectrum of the ions was obtained without the direct limitation of a limited mesh grid resolution. Ideally, if one aims to reconstruct measurements by a particle detector, the radius of the "particle cloud" at each trajectory point should be determined by the thermal velocity of the particles and the measurement or exposure time of the instrument. Notably, the larger the radius, the higher the number of particles within the cloud, resulting in improved statistical accuracy. We chose a radius of 5 ion inertial lengths in the solar wind (roughly 200 km) in order to obtain sufficient data points (N > 1000 per trajectory point) to compute a spectrum. 

The resulting ion kinetic energy spectrum along the trajectory is presented in Fig. \ref{fig:ion_profile}. It shows the particle distribution functions as histograms at each trajectory point alongside one another. The displayed energy ranges from 1 eV to 20 keV, with 60 logarithmic energy bins per histogram. We note that while Fig. \ref{fig:ion_profile} uses time as the X-axis, the analysis is based on adjusting the spatial position within a quasi-stationary state at the end of the simulation. To guide the eye, we indicate the solar wind initialization temperature of 1 keV with a horizontal dashed line in Fig. \ref{fig:ion_profile}. \\

During MSB2, the BepiColombo spacecraft entered the magnetosphere at the post-terminator duskside flank of the magnetosheath. Due to the size of the simulation box used, the spectrum in Fig. \ref{fig:ion_profile} panel a starts at 08:45 UTC, and the spacecraft is already located in the magnetosheath. The particle energy peaks are just under 1 keV, which equates to the solar wind input energy. Due to the distance of X > +4 $\mathrm{R_M}$ at Y $\sim$ -3 $\mathrm{R_M}$, the solar wind that was decelerated and heated at the bow shock accelerates again to solar wind speed \citep{gershman2013}. 

Nevertheless, considering the width of the energy distribution at about 09:10 UTC in Fig. \ref{fig:ion_profile} panel a, a few particles have energies that exceed 1 keV due to acceleration mechanisms at the duskside flank of the inbound magnetopause. After (1) in Fig. \ref{fig:ion_profile}, the energy decreases gradually to about 50 eV at 09:33 UTC. This lowering of the energy by one order of magnitude is seen as being a result of a passage from the magnetopause through the tail plasma lobes. Between 09:35 UTC and 09:40 UTC, marked with (2), a particle population with an energy range of about 20 to 200 eV is observed. Taking into account Fig. \ref{fig:ion_density} panels c and d, this is interpreted as a grazing of the tail current sheet from its southern boundary. In the area between (3) and (4), the density increases, and there is a particle population with energies of a few electron volts up to a maximum of 1 keV. 

Starting from (3) in Fig. \ref{fig:ion_profile}, the energy first decreases until about 09:47 UTC and then increases again until (4). This population can be identified with the trapped particles seen in Fig. \ref{fig:ion_density} panel a. These particles have a predominantly negative x-velocity component (see Fig. \ref{fig:ion_profile} panel b), meaning they move in a sunward direction, and compared to Fig. \ref{fig:ion_density} panel a, a convection pattern can be identified. These accelerated particles are due to magnetotail reconnection and drift planetward and toward the magnetopause. Between the closest approach and (4) there is another particle population accelerated to higher energies centered at about 6 keV and reaching up to 10 keV. These particles are accelerated near the (dawnside) flanks of the tilted tail current sheet. Due to the proximity to the downtail reconnection site, the energies are higher compared to the generally lower energy population between (3) and (4). Point (4) roughly marks the outbound magnetopause crossing. In the dayside magnetosheath, ranging from (4) to (5), the energy ranges from a 10 eV up to about 4 keV. At around 09:54 UTC, point (5), the spacecraft traversed through the bow shock into the solar wind. Between (5) and about 10:05 UTC, there are also ions with a sunward velocity component, despite the solar wind bulk flow (see Fig. \ref{fig:ion_profile} panel b), that potentially describe a field-aligned ion beam in the foreshock region that was identified in Fig. \ref{fig:ion_density} panels a and c.

While precise times are indicated, we note that an averaging effect exists in the energy spectrum as a result of employing a radius of about 200 km for the computation of the particle spectrum. Consequently, boundary regions become "blurred" and timings may not be exact. Moreover, the indicated particle energies may deviate from the real-world scenario, especially the low-energy population in the magnetopshere. In that regard, our simulation results are more appropriately viewed as a qualitative representation of the real flyby.

\section{Summary and conclusion}
We used a global 3D hybrid plasma simulation to investigate the entry of solar wind protons into Mercury’s magnetosphere. Particle densities as well as the electric current were used to evaluate magnetopsheric regions and boundaries. We utilized solar wind upstream IMF conditions derived from averaged measurements obtained by the BepiColombo MGF instrument during the MSB2 phase that are also commonly observed near Mercury \citep{james2017}. The MSB2 trajectory was used as an example to evaluate proton energy distributions throughout the southern magnetosphere and to quantitatively estimate boundary crossings.

The magnetopause and bow shock stand-off distances at the subsolar point are $\mathrm{R_{MP}}$ = 1.45 $\mathrm{R_M}$ and $\mathrm{R_{BS}}$ = 1.85 $\mathrm{R_M}$. The magnetopause distance falls within the range of anticipated values \citep{winslow2013}, and likewise, the bow shock distance corresponds well with values documented by \cite{slavin2009, winslow2013}. Looking at the general shape of the magnetosphere, a clear difference between empirical models and our simulation can be seen (Fig. \ref{fig:ion_density}). The asymmetry in the equatorial plane has been found to be a result of the oblique IMF direction during MSB2. Solar wind particles enter the magnetopshere as they pass through the bow shock and are deflected around the magnetopause. They follow magnetic field lines at the cusps toward the surface and can also enter the magnetopshere through magnetic reconnection at the magnetotail. During these reconnection processes, the particles are accelerated and convect around the planet toward the dayside magnetopause.

We initialized the simulation with solar wind protons that exhibit an average kinetic energy of 1 keV. The resulting energies of these protons influenced by Mercury's magnetosphere vary by three orders of magnitude, ranging from a few electron volts near the dawnside terminator lower than a few 100 km surface altitude (around the closest approach of MSB2) up to about 4 keV in the magnetosheath region on the pre-terminator dawnside, which is consistent with earlier observations by MESSENGER \citep{zhao2020}. Moreover we observed a grazing of the magnetotail plasma sheet. After the bow shock crossing, there are indications for an ion beam in the foreshock region of Mercury. 
 
For this study, we utilized a steady state snapshot of Mercury's magnetosphere. While the positions of boundaries and energy distributions were investigated and estimated, the dynamics of the system are not covered. Mercury's magnetospheric system, however, is constantly changing due to IMF variability \citep{fatemi2018, exner2018} and the fast Dungey cycle of about 2 minutes \cite{slavin2009}. Therefore, our results should be regarded as a quantitative representation, especially if compared to particle measurements during MSB2. The flyby interval of about one to two hours might allow reconfigurations that potentially affect boundary positions and orientations.

Future work will entail comparing modeled data with ion measurements. This could include investigating dynamic effects, methods to obtain directional particle spectra, and incorporating ion species of planetary origin to investigate exospheric processes. Moreover, simulations with varying IMF and solar wind conditions can be used to provide forecast scenarios for all upcoming Mercury swingbys, contributing to mission/operation planning of BepiColombo.



\begin{acknowledgements}
      This work was financially supported by the Austrian
      Research Promotion Agency (FFG) ASAP PICAM-4 under contract 885349. DT acknowledges the Mio MGF team (W. Baumjohann as PI, A. Matsuoka as Co-PI) for providing the mean values of interplanetary magnetic field for the BepiColombo flyby-2 outbound trajectory. Furthermore, DT acknowledges constructive discussions with colleagues from the Space Research Insitute (IWF) Graz. WE has been supported by an ESA research fellowship. WE and UM acknowledge the North-German Supercomputing Alliance (HLRN) for providing HPC resources that have contributed to the research results reported in this paper. VisIt 3.2.2 has been used to generate the 2D-plots \citep{Childs2012_VisIt}.
\end{acknowledgements}

%
%

\bibliographystyle{aa}
\bibliography{main}

\end{document}